\documentclass{aa} 
 
\usepackage{graphicx} 
\usepackage{txfonts} 

\begin{document} 
 
\title{Detecting companions to extrasolar planets using mutual events}
 
 
\author{J. Cabrera  
        \and  
        J. Schneider} 
 
 
\institute{Observatoire de Paris-Meudon, 92195, Meudon Cedex, France\\
           \email{juan.cabrera@obspm.fr} }
 
\date{Received ; accepted } 
 
\abstract 
{} 
{We investigate a new approach to the detection of companions to  
  extrasolar planets beyond the transit method. We discuss the
  possibility of the existence of ``binary planets''.} 
{We develop a method based on the imaging of a planet-companion as an  
  unresolved system (but resolved from its parent star). It makes use  
  of planet-companion ``mutual phenomena'', namely mutual transits 
  and mutual shadows.} 
{We show that companions can be detected and their radius measured  
  down to lunar sizes.}
{} 
 
 
\titlerunning{Detection of companions to exoplanets} 
 
\maketitle 
 
%
\section{Introduction} 
 
The search for satellites of extrasolar planets is relevant to the 
understanding of the evolution of planetary systems and to the
perspective of their habitability. Their occurrence in other planetary
systems is very likely since in the Solar System, 7 of the 9 ''planets''
have from 1 to several tens of satellites and numerical simulations of
planet formation show that such satellites should be common (Ida et 
al. \cite{ida97}). Several methods have been proposed 
for their detection: Sartoretti \& Schneider (\cite{sartoretti99})
proposed detection by transits; Brown et al. (\cite{brown01})
applied this approach to constrain companions to  
HD 209458 b; Han \& Han (\cite{han02}) and Bennett \& Rhie
(\cite{bennett02}) discussed the possibility of detecting
companions via microlensing; Williams \& Knacke (\cite{williams04})
have shown that Earth-like satellites of giant planets in the
habitable zone  would be detectable by spectroscopy in the
$\mathrm{CH}_4$ hole of the planet spectrum in the $1.5 - 4 \mu$m
band. Here we investigate new ways to detect planet satellites based
on planet wobble and photometry in direct imaging. First, we discuss
the possibility of binary planets and their impact on planet
characterization. 
 
%
 
In addition to satellites defined as companions with masses and sizes  
significantly smaller than their parent planets, we suggest that there  
may exist binary planets with comparable masses and sizes. Such binary
planets do not exist in the Solar System. However, the  
detection of the first 180 other planetary systems has shown much  
diversity and differences with the Solar System in their 
characteristics and provided surprises such as very small and highly  
eccentric orbits. These unexpected findings open the possibility of  
new configurations such as binary planets. In addition, binarity is  
not an exception in objects conceptually associated with planets,
such as asteroids (Pravec et al. \cite{pravec06}), trans-Neptunian
objects (Kern \& Elliot \cite{kern06}) or brown dwarfs (Stassun,
Matthieu \& Valenti \cite{stassun06}). In the same way, refined
simulations show the formation of vortices close to planets (Kley
\cite{kley03}) which could in turn form a massive companion to the
planet. 
 
An interesting counterargument has recently been put forward (Canup 
\& Ward \cite{canup06}) predicting that the mass ratio of the companion 
to its host planet cannot be larger than about $7~10^{-4}$. It will be 
interesting to see in the coming decade if this prediction is 
confirmed by observations. We note that this prediction does not hold 
for non giant Solar System bodies like the Earth and Pluto. 
 
%
 
The presence of companions to planets would have a significant  
impact on their characterization. 
 
\begin{itemize} 
\item  {\it Planet mass} 
 
For binary planets the radial velocity method of detection (RV) gives  
only the sum $M=M_1+M_2$ of their (minimum) masses, leading to a  
false assignment of the mass of individual objects. In particular if 
a companion detected by RV has a (minimum) mass larger than the 
standard planet mass upper limit ($\sim 13$ Jup. mass), it would be 
inappropriately discarded as being a brown dwarf and not a planet. The 
same consideration holds for the astrometric detection of planets. 
 
\item {\it Radius and albedo} 
 
The future detection of planets by direct imaging will give, in the
case of a planet-companion system, only the sum $F=F_1+F_2$ of their 
fluxes: \\ 
$(F_{\mathrm{Refl}})_{1,2} = A_{1,2} \times R_{1,2}^2$ 
(modulated by an orbital phase) for the reflected flux and\\ 
$(F_{\mathrm{Th}})_{1,2}=R_{1,2}^2 \times T_{1,2}^4$ (constant along the orbit) 
for the thermal emission. \\ 
The albedo $A$ of a cold planet cannot be larger than 1; its radius
cannot be larger than $\sim 1.1~R_{\mathrm{Jup}}$ (Guillot \cite{guillot05}). 
Therefore the normalized flux of a planet cannot be larger than the
maximum normalized flux:\\ 
$F_{\mathrm{max}} =     1.2~R_{\mathrm{Jup}}^2$ and  
$F_{\mathrm{max}} = T^4 1.2~R_{\mathrm{Jup}}^2$ \\ 
for the reflected and thermal flux respectively.

If the observed normalized flux of the planet is larger than the 
maximum possible flux $F_{\mathrm{max}}$, either there is something wrong with 
its structure or it is binary. An example of an odd structure is given 
by a planet surrounded by rings, since then there is an additional 
contribution to the planet flux coming from the planetary ring 
(Schneider \cite{schneider03}). Their presence can be detected in 
reflected light by a non-Keplerian modulation of their reflected flux 
along the orbit (Arnold \& Schneider \cite{arnold04}, 
Barnes \& Fortney \cite{barnes04}); in the
thermal regime there is no such orbital modulation and there is a
priori no way to disentangle a binary planet from a ringed planet.

If the observed planet normalized flux is smaller than $F_{\mathrm{max}}$ and 
if it is in reality a planet-companion system, the assumption that it 
is a single planet would lead to incorrect assignments to the albedo 
and/or radius of its components. 
 
\item {\it Planet spectrum} 
 
If there is a satellite with a significant contribution to the planet  
flux, the global planet + satellite spectrum may lead to 
misinterpretations if the spectrum is attributed to a single object  
(leading to incorrect atmosphere models). An important example is  
provided by an icy satellite of a telluric planet close to the  
habitable zone. The planet may undergo a greenhouse effect leading to
an atmosphere with water and CO2, while the satellite can be icy. In
that case, photo\-dissociation of the satellite's ice can lead to
ozone synthesis (Teolis et al. \cite{teolis06}); the global spectrum
would mimic an Earth spectrum.  
 
\end{itemize} 
 
The case of planets detected by transits is different. Their possible 
binarity would be easily inferred from the peculiar shape of the 
transit lightcurve or from their timing revealing non periodic transits 
(Sartoretti \& Schneider \cite{sartoretti99}). For the 9 planets 
detected by transit by 22 Aug. 2005 (HD 209458 b, HD 149026 b, 
TRES-1, OGLE-TR-10, OGLE-TR-111, OGLE-TR-113, OGLE-TR-132, OGLE-TR-56, 
HD 189733; see \mbox{\texttt{exoplanet.eu}}), none of them
presents any sign of a moon or of binarity. But their orbital radius
is so small ($a < 0.05\mathrm{AU}$) that 
the Hill radius $a(M_{\mathrm{pl}}/3 M_*)^{1/3}$ (Valtonen
\cite{valtonen06}) inside which a stable orbit can survive around the
planet would make these planets merge together (for a study of the
stability of satellites around giant extrasolar planets, see Barnes \&
O'Brien \cite{barnes02}). Binarity among ``hot Jupiters'' is 
therefore very unlikely. Only planets at orbital distances larger 
than $\sim 0.1 \mathrm{AU}$ can, from a dynamical stability point of view, be 
binary. But the probability that they make a transit drops as $1/a$ 
and only large surveys like the CoRoT and Kepler space missions will 
detect them. Finally, the detection by transits of a binary planet 
lacks generality since it requires a transit to occur. 
 
Thus, the detection of planets by RV, 
astrometry and direct imaging miss their possible binarity and could 
lead to misassignments of their mass, albedo and radius. The most 
radical method to detected the binarity of a planet would consist of 
resolving the planet-companion system in high angular resolution 
direct imaging. But, for a typical Jupiter/Ganymede system at 5pc, the 
baseline $B$ required to resolve the system at $2\mu$m would be $B=1.2 
\times 2\mu\mathrm{m} \times 5\mathrm{pc}/10^6\mathrm{km} =
360\mathrm{m}$.  In addition, a very high contrast of at least $10^6$
is required. Such a baseline and high contrast only will be achieved
in the future, here we investigate less difficult methods.  
 
%
\section{New approaches to planet-companion system detection} 
 
We consider the detection of a planet-companion system by 
direct imaging in which the system is unresolved and thus appears as a 
single point, supposed to be detached from its parent star. Gaidos et 
al. (\cite{gaidos06}) have considered the thermal emission
approach. Here we concentrate on the reflected light approach of
planet imaging.  
 
%
\subsection{Planet wobble}

 
Let $F_{\mathrm{pl}}= \pi A_{\mathrm{pl}} R_{\mathrm{pl}}^2$ and 
$F_{\mathrm{c}} = \pi A_{\mathrm{c}} R_{\mathrm{c}}^2$ be the flux of
the two components of the planet-companion system. The photocenter
will then make a wobble around the center of mass with an amplitude
$\Delta a$ given by  
 
\begin{equation} 
\Delta a = \frac{a_{\mathrm{c}}} 
                { M_{\mathrm{pl}} + M_{\mathrm{c }} } 
\left( \frac{ F_{\mathrm{pl}} M_{\mathrm{c }} - 
              F_{\mathrm{c }} M_{\mathrm{pl}} }
            { F_{\mathrm{pl}} + F_{\mathrm{c}}} 
\right).  
\end{equation} 
 
For a Saturn-Titan system, the mass and flux ratios are 
$M_{\mathrm{c}}/M_{\mathrm{pl}} = 0.01$ and 
$F_{\mathrm{c}}/F_{\mathrm{pl}} = 0.001$; then at 5 pc, the amplitude
of the angular wobble is 15 microarcsec. For a Jupiter-Saturn binary
system with the same separation 
$M_{\mathrm{c}}/M_{\mathrm{pl}} = 0.3$ and 
$F_{\mathrm{c}}/F_{\mathrm{pl}} = 1$; the angular wobble amplitude is
then 1 mas. For a perfect twin planet (same mass, radius
and albedo), the photocenter remains fixed along its revolution and
the binarity is undetectable by this approach. 
 
These numbers should be compared to the accuracy $\delta \theta$  
on the planet position from ground and space observations. This accuracy 
is given, at best, at a wavelength $\lambda$, for a bandwidth $W$ and 
a telescope of diameter $d$ by: 
 
\begin{equation} 
\delta \theta = \frac{1}{\sqrt{N_{ph}}}\frac{\lambda}{d} 
\end{equation} 
 
where $N_{\mathrm{ph}} = 10^3 \, 10^{-0.4 \, m} \, (d / 1 \, 
\mathrm{cm})^2 \, (W / 1 \, \AA) \, (T_{\mathrm{exp}}/1 \, 
\mathrm{sec})$ is the number of detected photons emitted by the
planet. For a \mbox{$ m = 25$} planet, 
$\delta \theta = 15 \, \mu \mathrm{as}$ (resp. $1 \, \mu \mathrm{as}$) 
in a \mbox{10 h} exposure with \mbox{$W = 100 \, \mathrm{nm}$} 
and \mbox{$d = 8 \, \mathrm{m}$} 
(resp. \mbox{$d = 30 \, \mathrm{m}$}) 
at $\lambda = 1 \, \mu \mathrm{m}$.  
 
Of course, any intrinsic flux variation of the planet and/or companion
would complicate the detection of the signal.

The planet-companion system can be seen as a spec\-tro\-scop\-ic 
bi\-na\-ry. Each component has a radial velocity variation with an 
amplitude $V_{1,2} = \left[ M_{2,1}/(M_1 + M_2 ) \right] 
\left( G \, (M_1 + M_2) / a_{\mathrm{c}} \right) ^{1/2}$. 
 
For a Saturn-Titan system, the exo-Saturn velocity is \mbox{2 m/sec}. For a 
Jupiter-Saturn binary system with the same separation the exo-Jupiter 
velocity is \mbox{10 km/sec}.  
 
The feasibility of this wobble approach has been addressed in the 
context of Extremely Large Telescopes by Ardeberg (2005). 
 
%
\subsection{Mutual photometric phenomena} 
 
\begin{figure} 
  \resizebox{\hsize}{!}{\includegraphics{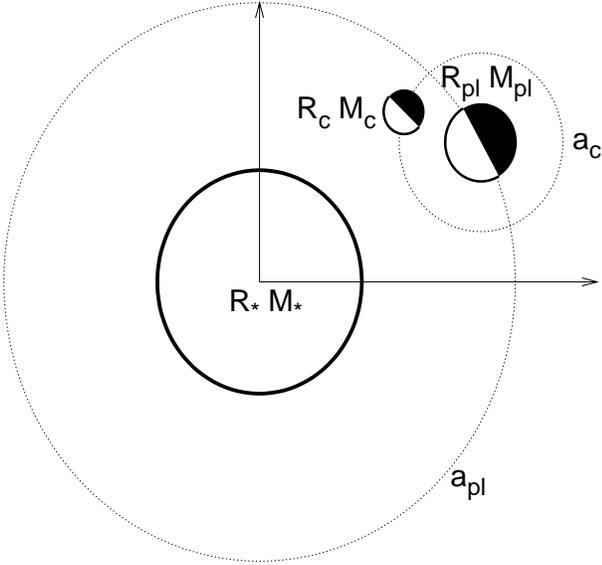}} 
  \caption{Schematic view of the system (not to scale). The
  companion's orbit around the planet and planet's orbit around the
  star do not need to be coplanar. The white surfaces of the planet
  and the companion are illuminated, while black surfaces are in
  darkness.}
  \label{FigSchema} 
\end{figure}

Let us consider a restricted three body problem consisting of a star, a
planet orbiting the star and a companion orbiting the planet. In our
problem, the companion (with radius $R_{\mathrm{c}}$ and mass
$M_{\mathrm{c}}$) and the planet (with radius $R_{\mathrm{pl}}$ and mass
$M_{\mathrm{pl}}$) orbit each other in a circular orbit of semimajor
axis $a_{\mathrm{c}}$; the center of mass of this system orbits the
star (with radius $R_{\mathrm{*}}$ and mass $M_{\mathrm{*}}$) with a
semimajor axis of $a_{\mathrm{pl}}$ (see Fig. \ref{FigSchema}).

The only source of light is the star, which shines with constant
luminosity $P$ (following the notation of Lester et
al. \cite{lester79}). The flux (density) $F_{*}$ arriving at the planet
and the companion is constant and uniform. The flux reflected by these
bodies is not resolved, but it is detached from the light coming from
the star. This reflected flux depends on the physical characteristics
of the reflecting surfaces and the geometrical configuration of the
system.

The theory that describes the radiation by a planet illuminated by its 
star is well described in Lester et al. (\cite{lester79}) and Fairbairn 
(\cite{fairbairn02}) for example. Here we will suppose that both the 
planet and the satellite are Lambertian spheres. Scattering 
atmospheres may differ from this model but ours is still a good
approximation.

In these circumstances, the flux reflected by the planet (see
Eq. \ref{EqflplEarth}) is proportional to:

\begin{itemize}
\item its geometrical albedo $p$;
\item its surface (that is, the square of the radius $R$);
\item the inverse of the square of its distance to the star $a$;
\item the \emph{phase law} $\Psi \left( \alpha \right)$.
\end{itemize}

\begin{figure} 
  \resizebox{\hsize}{!}{\includegraphics{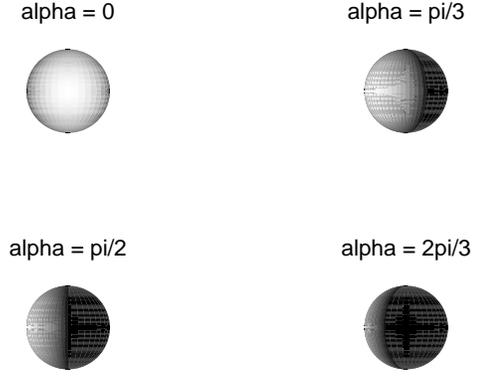}} 
  \caption{Illumination of the planet as seen from the observer for 
  different phase angles.} 
  \label{FigPhaseLaw} 
\end{figure}

The phase law is an integral which takes into account the geometry of
the system. It depends on the \emph{phase angle} $\alpha$, which is
the angle between the incident direction of the radiation and the
direction to the observer, as seen from the center of the sphere (see
Fig. \ref{FigPhaseLaw}). The integral has the following form: 
 
\begin{equation} \label{Eqphlw} 
\Psi \left( \alpha \right) = \frac{3}{2 \pi} \int\!\!\!\int d\theta \, 
d\phi \, \sin^{3}(\theta) \, \cos(\phi) \, \cos(\phi + \alpha).  
\end{equation} 

With these assumptions, the flux reflected by the planet as seen from
Earth is: 
 
\begin{equation} \label{EqflplEarth} 
F_{\mathrm{pl}} = \mathrm{p} \, \frac{ R_{\mathrm{pl}}^{2} }{
  a_{\mathrm{pl}}^{2} } \, \Psi \left( \alpha \right) \, F_{*} 
\end{equation} 

The limits of integration of Eq. (\ref{Eqphlw}) are $\theta \in [0,\pi]$ and  
$\phi \in [-\pi / 2 , \pi / 2 + \alpha ]$; so it takes the form: 

\begin{equation} \label{Eqphlwint} 
\Psi \left( \alpha \right) = \frac{ \sin( \alpha )+( \pi - \alpha ) 
  \cos( \alpha ) }{\pi}. 
\end{equation} 

For further details on how to arrive at these expressions, please
see Lester et al. (\cite{lester79}).

\begin{figure} 
  \resizebox{\hsize}{!}{\includegraphics{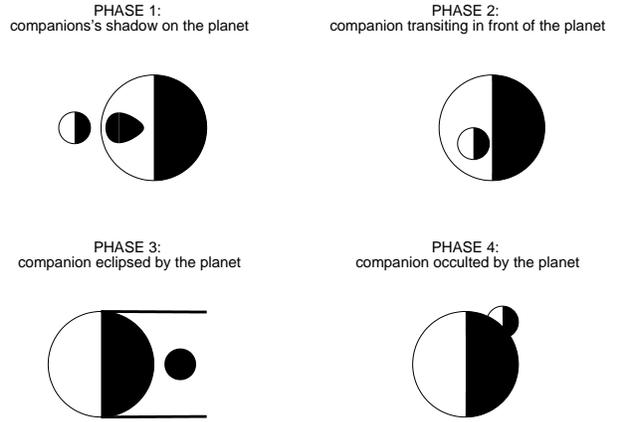}} 
  \caption{Mutual phenomena during the revolution of a companion 
  around its parent planet; seen from the point of view of the
  observer; illumination comes from the left.} 
  \label{FigFases} 
\end{figure}

In general, the total flux that we measure is the addition of
the individual fluxes coming from each body. However, if the orbit of
the satellite lies on the line of sight of the observer, the orbital
revolution of the satellite around the planet has 4 remarkable phases
(see Fig. \ref{FigFases}): 
 
\begin{itemize} 
\item[-] Phase 1: companion's shadow on the planet. 
\item[-] Phase 2: companion transiting in front of the planet. 
\item[-] Phase 3: companion eclipsed by the planet.  
\item[-] Phase 4: companion passing behind the planet and being
  occulted by it.
\end{itemize} 
 
In any of these cases we see a flux decrease because part of the
reflecting surface is in darkness. 

\begin{figure} 
  \resizebox{\hsize}{!}{\includegraphics{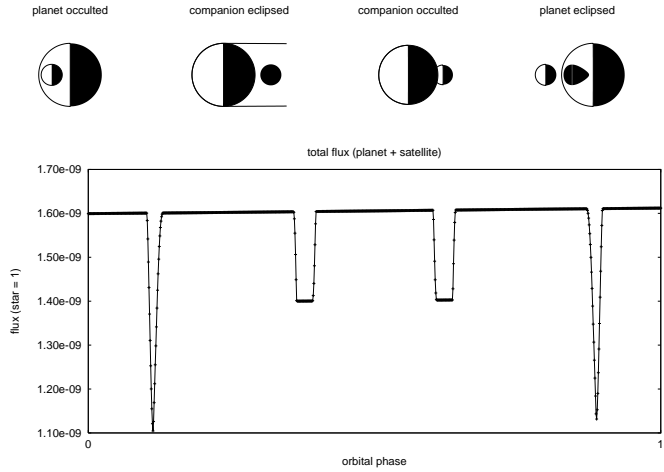}} 
  \caption{Lightcurve for a planet with a companion; the phase angle
  is $\alpha = 90^{\circ}$; $R_{\mathrm{c}}/R_{\mathrm{pl}} = 1/3$;
  the trend in the out-of-event part of the curve is due to the
  variation in phase of the planet, as it orbits the star.}
  \label{graficas01} 
\end{figure}

\begin{figure} 
  \resizebox{\hsize}{!}{\includegraphics{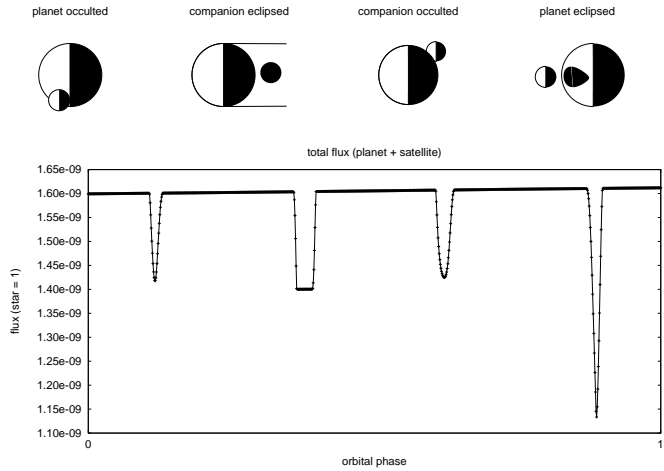}} 
  \caption{Lightcurve for a planet with a companion; the phase angle
  is $\alpha = 90^{\circ}$; $R_{\mathrm{c}}/R_{\mathrm{pl}} = 1/3$; the
  orbit of the companion is inclined $85^{\circ}$ with respect to the
  line of sight.}
  \label{graficas02} 
\end{figure}

\begin{figure} 
  \resizebox{\hsize}{!}{\includegraphics{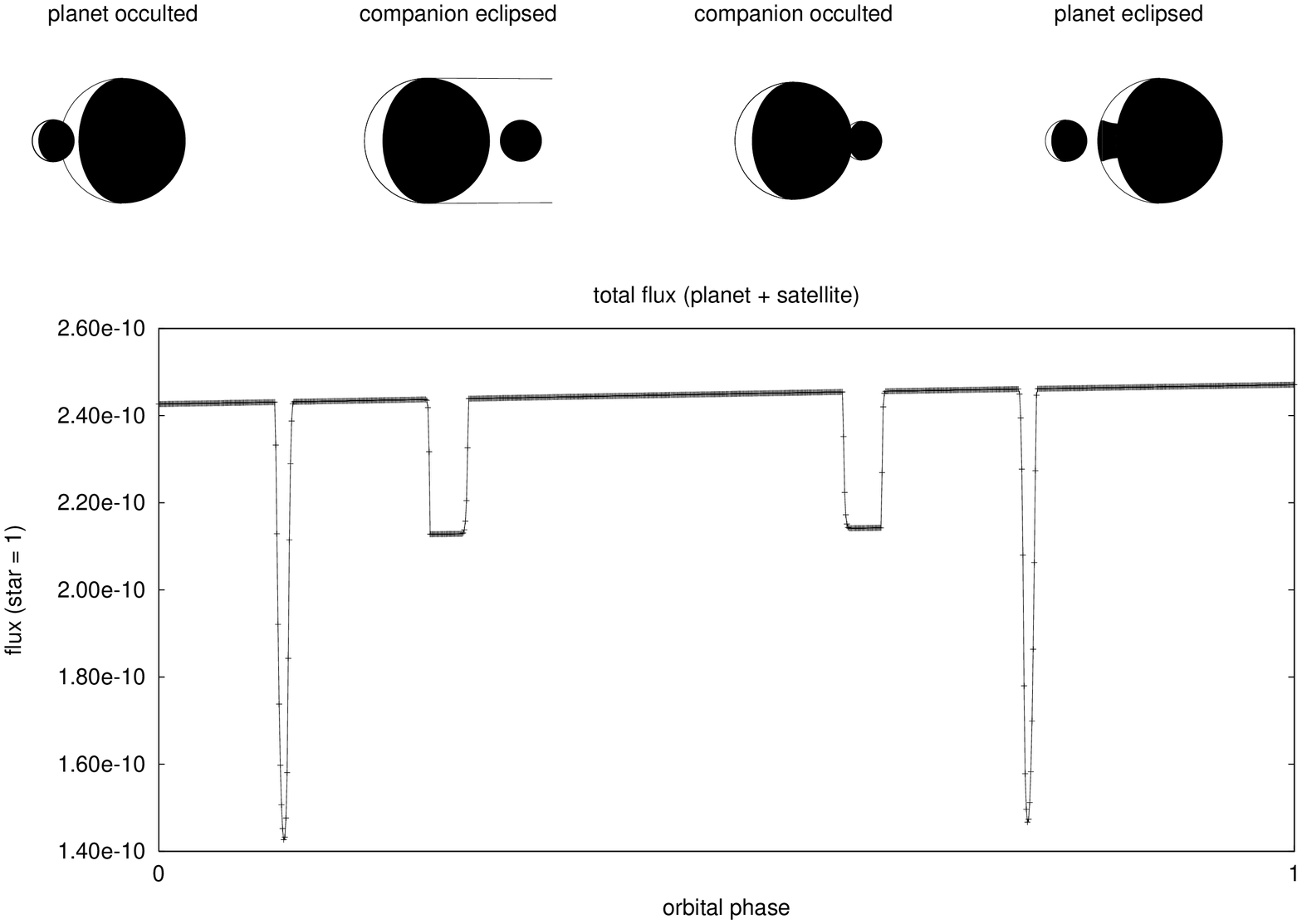}} 
  \caption{Lightcurve for a planet with a companion; the phase angle
  is $\alpha = 135^{\circ}$; $R_{\mathrm{c}}/R_{\mathrm{pl}} = 1/3$}
  \label{graficas03} 
\end{figure}

\begin{figure} 
  \resizebox{\hsize}{!}{\includegraphics{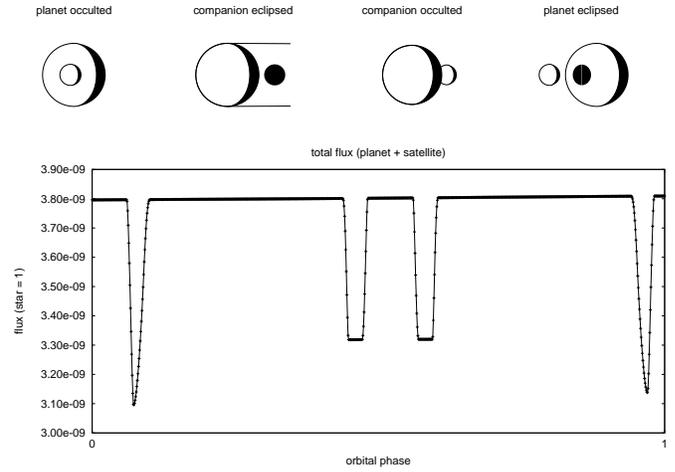}} 
  \caption{Lightcurves for a planet with a companion; the phase angle
  is $\alpha = 45^{\circ}$; $R_{\mathrm{c}}/R_{\mathrm{pl}} = 1/3$}
  \label{graficas04} 
\end{figure}

To calculate this flux decrease we consider the surface in darkness and we
integrate Eq. (\ref{EqflplEarth}) within the limits of this
surface. We have calculated this flux numerically and we present it in
Figs. \ref{graficas01} to \ref{graficas04} for different
configurations of the system. In the next subsections we will detail
these calculations. The signal in the
lightcurves is not periodic any more (as it was in the transits of a
single planet in front of a star); the position, shape and
depth of the flux decrease strongly depend on the relative positions of
all the bodies involved (star, planet, companion and observer).

\begin{enumerate}
\item[] {\bf Mutual transits}
\end{enumerate}

One of the bodies is transiting in front of the other, occulting its
surface. This surface is the intersection of a cylinder with a
sphere: a classic problem of surface geometry. However, the
expression resulting from integration (\ref{EqflplEarth}) within the
limits imposed by the intersection is not analytic. In the next two
points we will approximate that integral for the two possible
transits, in order of difficulty.

The geometric probability of transits is 
$R_{\mathrm{pl} } / a_{\mathrm{c}} \approx 10 \% $.
 
\begin{enumerate} 
\item[-] {\bf Phase 4: companion occulted by the planet}  
\end{enumerate} 
 
The companion is supposed to be smaller than the planet, so except in
the ingress and egress phases, the companion is completely
occulted\footnote{if the plane of the orbit of the satellite is
  inclined, this may not be true (see Fig. \ref{graficas02}); in that
  case we will have to calculate which is exactly the occulted
  surface of the companion.}. Using Eq. (\ref{EqflplEarth}), the
corresponding flux decrease is: 
 
\begin{displaymath} 
\frac{\Delta F_{\mathrm{T}}}{F_{\mathrm{T}}} = 
\frac{F_{\mathrm{c}}}{F_{\mathrm{pl}}+F_{\mathrm{c}}} = 
\frac{p_{\mathrm{c}}  \, R_{\mathrm{c}}^2}
     {p_{\mathrm{pl}} \, R_{\mathrm{pl}}^2 + 
      p_{\mathrm{c}}  \, R_{\mathrm{c}}^2}.
\end{displaymath} 
 
The duration of this transit is:

\begin{displaymath} 
\Delta T_{4} = \frac{ P_{\mathrm{c}} \, R_{\mathrm{pl}} }
                    { \pi \, a_{\mathrm{c}} }.
\end{displaymath}
 
\begin{enumerate} 
\item[-] {\bf Phase 2: companion transiting in front of the planet}  
\end{enumerate} 
 
We can consider the maximal occultation: this is, the maximum
(or minimum, depending on the reader's choice) of the flux
decrease. To do so, we suppose that the surface
occulted\footnote{regarding 
  Eq. (\ref{Eqphlw}) we can easily verify that the coordinates of the
  brightest point of the surface are $\theta = \pi / 2$ and $\phi = -
  \alpha / 2$.} has a constant brightness (which is not true) and we 
approximate this area by the surface of a disk with the companion's
radius. The corresponding flux decrease is: 
 
\begin{displaymath} 
\frac{\Delta F_{\mathrm{T}}}{F_{\mathrm{T}}} \approx  
\frac{ p_{\mathrm{pl}} \, R_{\mathrm{c}}^{2} \, \frac{3}{2} \cos \alpha/2 }  
  { (  p_{\mathrm{pl}} \, R_{\mathrm{pl}}^{2} + 
       p_{\mathrm{c}}  \, R_{\mathrm{c}}^{2} ) \, \Psi( \alpha ) } 
\end{displaymath} 
 
and the event duration is:

\begin{displaymath} 
\Delta T_{2} = \frac { P_{\mathrm{c}} \, R_{\mathrm{pl}} \,
                                        (1 + \cos \alpha ) } 
                     { 2 \, \pi \, a_{\mathrm{c}} }.
\end{displaymath}
 
\begin{enumerate}
\item[] {\bf Mutual shadows}
\end{enumerate}
 
In phases 1 and 3 one body eclipses the other. In eclipses we have
to differentiate the \emph{umbra} (defined as the region of an eclipse where
the light coming from the primary source is completely excluded) from
the \emph{penumbra} (the region where only part of the light from the primary
source is excluded). However, in the case of planets orbiting at
distances similar to those of Jupiter or Saturn, the fraction of time
in which the companion is only in penumbra compared with the fraction
of time in which it is in umbra is of the order of $10^{-4}$ (whereas
for the Earth-Moon system it is 0.7); so we decided to neglect
it\footnote{The effect of penumbra will be to soften the sharpness of
  the egress and ingress, which are very difficult to observe anyway.}.

In this case the surface of interest is the intersection of a cone
with a sphere. In the next two points we will approximate the integral
(\ref{Eqphlw}) for the eclipses.

If the companion's orbital plane is close to the planet ecliptic, the
geometric probability of mutual shadows is close to 100\%. 
 
\begin{itemize} 
\item[-] {\bf Phase 3: companion eclipsed by the planet} 
\end{itemize} 
 
The companion disappears completely in the shadow of the parent
planet. The relative flux variation is then: 
 
\begin{displaymath} 
\frac{\Delta F_{\mathrm{T}}}{F_{\mathrm{T}}} = 
\frac{F_{\mathrm{c}}}{F_{\mathrm{pl}}+F_{\mathrm{c}}} = 
\frac{p_{\mathrm{c}}  \, R_{\mathrm{c}}^2}
     {p_{\mathrm{pl}} \, R_{\mathrm{pl}}^2 + 
      p_{\mathrm{c}}  \, R_{\mathrm{c}}^2}
\end{displaymath} 
 
with a duration of:

\begin{displaymath} 
\Delta T_{3} = \frac{ P_{\mathrm{c}} \, R'_{\mathrm{pl}} }
                    { \pi \, a_{\mathrm{c}} }
\end{displaymath} 
 
$R'_{\mathrm{pl}}$ is the radius of the perpendicular section of the
planet's shadow cone at the position of the companion:

\begin{displaymath} 
R'_{\mathrm{pl}} = 
R_{\mathrm{pl}} -
\frac{R_{*} - R_{\mathrm{pl}}} {a_{\mathrm{pl}}} a_{\mathrm{c}}  = 
R_{\mathrm{pl}} \left( 1 -
\frac{R_{*} - R_{\mathrm{pl}}} {R_{\mathrm{pl}}} 
\frac{a_{\mathrm{c}}} {a_{\mathrm{pl}}}
\right).
\end{displaymath} 

\begin{itemize} 
\item[-] {\bf Phase 1: satellite's shadow on the planet}  
\end{itemize} 
 
In this case the companion projects a shadow on the surface of the
planet. To calculate the flux decrease we proceed as in phase 2 and we
obtain:

\begin{displaymath} 
\frac{\Delta F_{\mathrm{T}}}{F_{\mathrm{T}}} \approx  
\frac{ p_{\mathrm{pl}}  \, R_{\mathrm{c}}^{'2} \, \frac{3}{2} \cos \alpha/2 }  
  { (  p_{\mathrm{pl}}  \, R_{\mathrm{pl}}^2 + 
       p_{\mathrm{c}}   \, R_{\mathrm{c}}^2 ) \, \Psi( \alpha ) } 
\end{displaymath} 
 
where $R'_{\mathrm{c}}$ is calculated in the same way as
$R'_{\mathrm{pl}}$ in \emph{phase 3}. The duration of the passage is:

\begin{displaymath} 
\Delta T_{1} = \frac { P_{\mathrm{c}} \, R_{\mathrm{pl}} \, 
                                              (1 + \cos \alpha ) } 
                     { 2 \, \pi \, a_{\mathrm{c}} }.
\end{displaymath}

The light curves presented here are free of noise, which
obviously would make the detection of these events more difficult.

\begin{itemize} 
\item[] {\bf Mutual phenomena in the thermal regime}  
\end{itemize} 

In the thermal regime, where one detects the thermal emission of a
planet and its companion, mutual transits occur in the same way as 
for reflected light and the corresponding light curves have the same
shapes. For shadows and eclipses the situation is different. The flux
decrease is related to the decrease in temperature of the part of the
planet and its companion not illuminated by the parent star. This drop
in temperature is sudden when the planet or companion surface is solid
(such as during the lunar eclipses). The flux decrease is then
identical to the reflected light case. When the planet or its
companion have a thick atmosphere, its thermal inertia and atmospheric
circulation inhibit the temperature drop and no flux decrease is
seen. For thin atmospheres the situation is intermediate. The amount
of thermal infrared flux decrease during shadows and eclipses thus
provides a way to estimate the thermal inertia of the surface of the 
planet and its companion (Spencer \cite{spencer87}).

%
\section{Feasibility and search strategy} 
 
\subsection{Photometric accuracy and detection limits} 
 
For a companion to be detectable, the photometric accuracy must be 
better than the transit or shadow decrease in an exposure time shorter
than, say, half the duration of the phenomenon. The durations are
typically of the order of 
\mbox{$P_{\mathrm{comp}}R_{pl}/\pi a_{\mathrm{comp}}$}, 
i.e. \mbox{8 h} for a Saturn-Titan system and \mbox{4 h} for an
Earth-Moon system. The photometric accuracy on the detection of a
planet is not controlled by its own photon noise, but by the photon
noise of the background consisting essentially, in real situations, of 
the speckles of the parent star halo. 

In current coronagraphic detection projects of planets in the visible,
the speckle background is typically 100 times the planet signal, at an
angular separation of 1.5 to 2 $\lambda /D$ ($D$ = telescope diameter)
for a \mbox{m = 5 star} and coronagraphic rejection factor of the star of
$50\,000$ on the stellar peak. This figure holds for Jupiter-like
planets at \mbox{$1\,\mathrm{AU}$} (planet to star flux ration
$10^{-8}$) detected by a 1.5 m telescope and for Earth-like 
planets at \mbox{$1\,\mathrm{AU}$} (planet to star flux ratio
$10^{-10}$) detected by a 7.5 m TPF-C. 

For a companion with one third the planet radius the relative flux
decrease during shadows and transits is \mbox{$\sim \! 10\%$}, i.e. 
$10^{-3}$ the speckle background. The detector must collect $10^7$
speckle + planet photons to detect the companion with a \mbox{SNR = 3}
in the speckle background. For a Jupiter-sized planet the accumulated
exposure time required is \mbox{30 h} for a \mbox{m = 5} star with a
\mbox{1.5 m} telescope. Since this exposure time exceeds the duration
of the event (\mbox{8 h} in case of a Saturn/Titan-like system) by a
factor of 4, the companion is detectable in a continuous planet 
monitoring only in cumulative exposures over 4 revolutions around the
planet, i.e. after 2 months of continuous monitoring. This is for
instance the type of monitoring planned in the 'Super-Earth Explorer'
project (Schneider et al. \cite{schneider06}). For TPF-C, the
collecting area is 20 times larger and correspondingly the exposure
time to detect a $0.3R_{\mathrm{Jup}}$ companion drops to 
\mbox{$30 \, \mathrm{h} / 20 = 1.5 \, \mathrm{h}$}. Note that a
continuous monitoring is also required for the detection of surface
inhomogeneities of planets (Ford et al. \cite{ford01}).

\subsection{Geometric probability of shadows and transits and 
  observation strategy}
 
The geometric probability of mutual transits is
\mbox{$p=R_{\mathrm{pl}}/a_{\mathrm{sat}}$}. For a Saturn/Titan-like
configuration it is approximately 10\%.  The probability that a
planet and its companion make mutual eclipses (with respect to the
parent star) depends on the inclination of the orbital plane of the
planet-companion system with respect to the orbit of the system around
the parent star. If both objects have been formed by accretion in a
protoplanetary disk, it it is likely that the two orbital planes are
close to each other. Thus mutual shadows should occur at any position
of the planet-companion orbit around the star. Consequently, the
geometric probability of mutual shadows should be nearly 100\%. 
 
For mutual shadows, which have the maximum geometric probability of 
occurrence, the total fraction of time for which the event occurs is: 
 
\begin{equation} 
\frac{\Delta T_1 + \Delta T_3}{P_{\mathrm{comp}}} \approx 
\frac{3}{\pi}\frac{R_{\mathrm{pl}}}{a_{\mathrm{comp}}}. 
\end{equation} 
 
It is of the order of 10\% for a Saturn-Titan system. It is thus 
necessary to have a duty cycle of at least 90\% in order not to miss 
this event. The duration of this high duty cycle must be at least 
$P_{\mathrm{comp}}$, i.e. 15 days for a Saturn-Titan system. In conclusion, the 
detection of companions by mutual shadows (the most probable event) 
requires continuous imaging of the planet for at least about two 
weeks. Such continuous imaging is possible only from space.

%
%

\end{document}